\pdfoutput=1
\documentclass[aps,prl,reprint,twocolumn,floatfix,showpacs]{revtex4-1}

\usepackage{graphicx}
\usepackage{amsmath,amssymb}
\usepackage{natbib}

\begin{document}

\title{Evolutionary optimization of optical antennas}
\author{Thorsten Feichtner}
\author{Oleg Selig}
\author{Markus Kiunke}
\author{Bert Hecht}
\affiliation{Nano-Optics \& Biophotonics Group, Department of Experimental Physics 5,\\R\"ontgen
Research Center for Complex Material Research (RCCM),\\Physics Institute,
University of W\"urzburg, Am Hubland, D-97074 W\"urzburg, Germany}

\begin{abstract}
The design of nano-antennas is so far mainly inspired by radio-frequency technology. However, material properties and experimental settings need to be reconsidered at optical frequencies, which entails the need for alternative optimal antenna designs. Here a checkerboard-type, initially random array of gold cubes is subjected to evolutionary optimization. To illustrate the power of the approach we demonstrate that by optimizing the near-field intensity enhancement the evolutionary algorithm finds a new antenna geometry, essentially a split-ring/two-wire antenna hybrid which surpasses by far the performance of a conventional gap antenna by shifting the n=1 split-ring resonance into the optical regime.
\end{abstract}

\maketitle

Light-matter interaction, i.e.~absorption and emission of light as well as the control of its spectral and directional properties, can be optimized by means of antenna-like plasmonic nano structures \cite{Bharadwaj2009, Biagioni2012}. This is of immediate importance in diverse  fields of research ranging from solar energy conversion \cite{Atwater2010}, photocatalytic \cite{Liu2011a} and sensing applications \cite{Becker2010} to single-particle manipulation \cite{Juan2009,Zhang2010} and spectroscopy \cite{Kinkhabwala2009} as well as quantum optics and communication \cite{Akimov2007,Kolesov2009,Huang2009,Jacob2011}.

RF-antenna designs are usually optimized for thin, infinitely good conducting wires that only support surface currents and are typically fed by transmission lines connected by infinitely narrow gaps \cite{Balanis1992}. For antennas at optical frequencies the general operation conditions deviate substantially from such ideal behaviour: (i) Antenna wire diameters are comparable to the electromagnetic penetration depth into the wire material leading to volume currents \cite{Dorfmueller2010}. In the case of noble metals, such wires therefore exhibit plasmon resonances in the visible spectral range resulting in a reduced effective wavelength of wire waves \cite{Novotny2007}. (ii) Feeding (excitation) of optical antennas is often achieved by focused laser beams or quantum emitters.  (iii) high-frequency-related effects such as the 'kinetic inductance' become significant \cite{Zhou2005}. It can therefore not be taken for granted that RF-inspired antenna designs, like dipole \cite{Muehlschlegel2005}, bow tie \cite{Schuck2005,Farahani2005} and Yagi-Uda antennas \cite{Taminiau2008,Curto2010}, represent 'optimal' geometries also at optical frequencies, although they provide a reasonable performance.

Evolutionary algorithms (EAs) find optimized solutions to highly complex non-analytic problems by creating subsequent generations of individuals coded by their respective genomes that compete for the right to pass on their properties, according to a fitness parameter \cite{Sivanandram2008}. These optimized solutions can then be analyzed to foster the understanding of underlying physical principles. Evolutionary optimization has successfully been applied in various fields of research, including pulse shape optimization in coherent control of chemical reactions \cite{Baumert1997} and field localization in plasmonic structures \cite{Aeschlimann2007,Aeschlimann2010}.  Furthermore, evolutionary optimization has been used to aid the development of  radio-wave antennas \cite{Huang2007,Pantoja2007}. First attempts to employ such methods to find improved plasmonic nanostructures have also been undertaken \cite{Ginzburg2011,Kessentini2011,Forestiere2010,Forestiere2012}, however, the investigated configurational space remained very limited.

Here we employ the method of evolutionary optimization in a more general setup to find improved plasmonic antenna structures that, in terms of near-field intensity enhancement (fitness parameter), outperform the best radio-wave-type reference antennas by a factor of two. Analysis of the fittest antenna reveals that it is a split-ring/two-wire antenna hybrid (Split-Ring-Antenna), which merges features of the fundamental magnetic resonance of a split ring with the fundamental electric resonance of a linear dipole antenna both in the visible wavelength regime.

\paragraph{Methods}

As fitness parameter we choose the normalized near-field intensity enhancement in the focus of an illuminating Gaussian beam ($\lambda_i=$647~nm, NA$=1$, 4.3 fs pulse duration, 144 nm bandwidth). A genetic representation of complex-shaped thin-film nano antennas is realized by composing structures (matrix antennas) from discrete gold cubes with fixed dimensions (10x10x11~nm$^3$) positioned on a 21$\times$21 square matrix in vacuum oriented perpendicular to and centered on the optical axis of the Gaussian beam. The fields of such structures can be described by local Maxwell equations.

\begin{figure}[htp]
\centering
\includegraphics[width=\columnwidth]{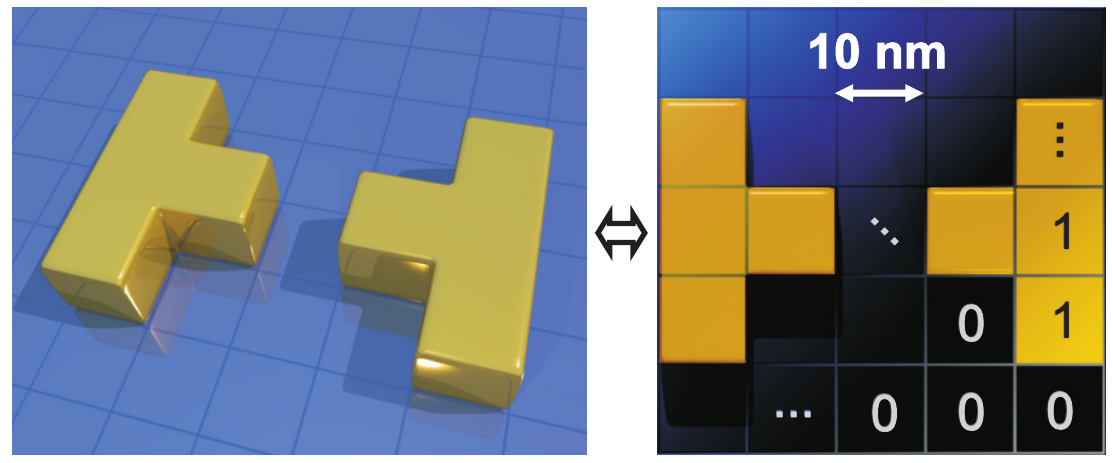}
 \caption{\label{fig:theory} (color online) Genetic representation of a matrix antenna. Left:~artistic 3D-view of an example bow-tie nano antenna consisting of 8 gold cubes. Right: top view indicating the transition to a 5$\times$5-matrix representation, where each '1' denotes the presence of a  10$\times$10$\times$11 nm$^3$ block of gold.}
\end{figure}

The resulting configuration space of about 4$\cdot 10^{132}$ different individual structures ensures geometrical variety but is impossible to explore by brute force methods, since the evaluation of an individual structure takes about 20 minutes. The size of the focal spot and the area occupied by the gold cube matrix are comparable (see Fig.~S1 \cite{suppLink}). An example showing a bow-tie antenna represented in a 5$\times$5 array is depicted in Fig. \ref{fig:theory}. The genetic information is represented in a unique binary code, where Matrix elements are set to '1', if occupied by a gold cube and to '0' when empty (for a calculation of the number of physical redundant structures see Fig.~S2, S3 and related discussion).

We solve Maxwell's equations using the finite-difference time-domain (FDTD) method \cite{Taflove2005} (FDTD Solution, Lumerical Inc., Canada). Due to the discretization in space (Yee Cells, dx = 1 nm), adjacent gold cubes are fully connected to each other. We also obtain conductive bridges between neighbouring gold blocks that contact each other via their edges. All simulation objects were shifted by half a Yee cell size in $x$- and $y$-direction to ensure identical dimensions of gold cubes and voids. In order to describe the dielectric function of gold in a sufficiently wide spectral range we use the data by Johnson and Christy fitted by an analytical model \cite{Etchegoin2006}.

The evolutionary algorithm is implemented in MatLab \footnote{All necessary code written in Lumerical script language or MatLab can be made available upon request.}. It uses generations consisting of 20 or 30 individual matrix antennas. The five best structures according to the fitness parameter are selected as parents for the next generation. With carefully chosen mechanisms for crossover and mutation of the genomes, consecutive generations constantly improve in the sense that their fitness parameter increases.  To create descendants in a first step one of the five parents is selected by roulette wheel selection \cite{Sivanandram2008} with a probability that is proportional to its fitness (parent $A$). Three methods, i.e. creation of random structures, mutation as well as linear and spiral genome crossing, are then applied until a total of 20 (or 30) new individuals have been generated for the next generation (see Fig.~S4).

After a sufficient amount of generations has been simulated, a so called \emph{toggle plot analysis} is performed, which consists of running 21x21 simulations in which every block is toggled individually. Colour-coding the block positions according to the magnitude of the fitness changes associated with the individual toggle event shows the relative importance of single blocks, the potential for further improvement of a matrix antenna, and also eventually produces new individuals with enhanced fitness \footnote{This is the brute force realization of the hill climber algorithm on our setup, neglecting pair- and higher-order correlations.}.

\paragraph{Results}

\begin{figure*}[htp]
\centering
\includegraphics[width=\textwidth]{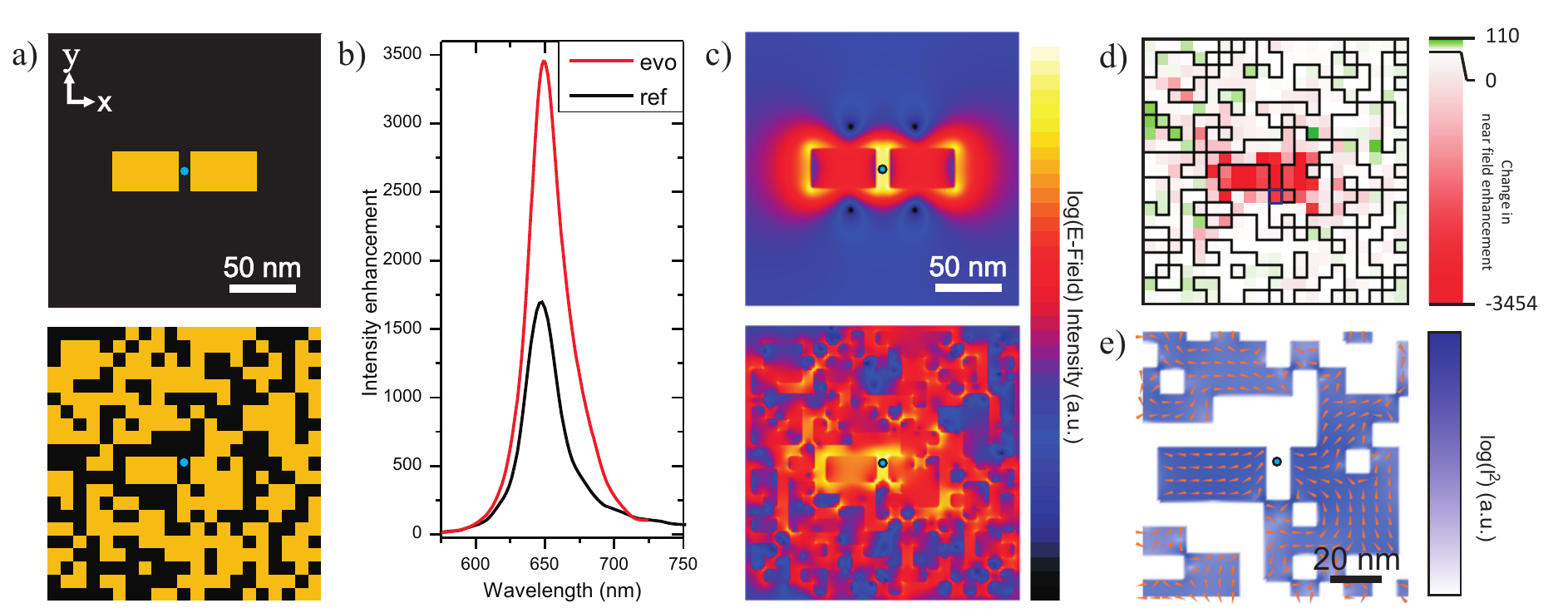}
\caption{\label{fig:comparison} (color online) Comparison at at $\lambda=$647 nm between a resonant linear dipole nano antenna build from two 10 nm separated rectangular arms of 46x30x11 nm$^3$ and a structure obtained with the evolutionary algorithm described in the text. (a) shows the geometry of both structures from the +$z$ direction. The blue spot denotes the position of the near field optimization by the EA. The spectra in (b) are taken at this marked position during a broadband gaussian excitation. (c) shows the  logarithmic near field intensities at $\lambda=$647 nm when the structures are illuminated by a monochromatic gaussian focus with NA = 1. The scales are normalized and not comparable. In (d) the the change of near field enhancement at the optimization position is shown for each single block, when it is toggled. (e) is a zoom of the centre part of the EA antenna, showing strength and direction of the currents.}
\end{figure*}

The wavelength $\lambda_i$ coincides with the resonance of a linear dipole nano antenna consisting of two end-to-end aligned 46$\times$30$\times$11 nm$^3$ gold rods (width = 3 cubes, height = 1 cube) separated by a 10 nm gap (Fig.~\ref{fig:comparison}(a), top panel) which serves as a reference structure. It exhibits a resonant normalized near-field intensity enhancement of about 1800 in the center of its feed gap. Other geometries, such as bow-tie antennas, were also tested but do not yield higher fitness.

In the following we discuss the fittest structure obtained by running the EA for 100 generations with 20 individuals each (see Fig.~S5 for development of geometry and fitness parameter), a subsequent toggle plot analysis and further 30 generations with 30 individuals each, starting with combinations of the best five structures obtained from the toggle plot analysis.

The best matrix antenna structure (Fig.~\ref{fig:comparison} (a), lower panel) exhibits a remarkably high fitness, as indicated by its near-field spectrum in Fig.~\ref{fig:comparison}(b) which is recorded in the optimization point (indicated as blue dot in (a)) after a broadband excitation. Its maximal near-field intensity enhancement of 3500 is nearly twice as high as that of the reference antenna. Both spectra show single nearly Lorentzian peaks (Q=20 and 23, respectively; see supplementary information for a further discussion of the Q-factor).

According to the reciprocity theorem \cite{Bharadwaj2009, Biagioni2012} the optimized antenna should also improve the radiative properties of a quantum emitter positioned in the spot of highest field enhancement. Indeed for the reference antenna we find a radiation enhancement of 2126 and a radiation efficiency of 0.255, while for the fittest antenna the radiation enhancement is 4271 with a radiation efficiency of 0.268. Surprisingly, the directivity of the fittest antenna remains very similar to that of the reference antenna despite its complex shape (see Fig. S6 and related discussion).

The fittest matrix antenna exhibits three noticeable geometrical features:

(i) a small gap in the center between two compact rod(-like) structures, being slightly displaced in y-direction with respect to the observation point.

(ii) a single gold block directly below the gap which creates a current path connecting the rod-like structures and 

(iii) a seemingly random arrangement of gold blocks further away from the center. 
\newline It is important to note that the optimal structure found by a genetic algorithm depends on both the available primitive elements as well as on possible boundary conditions. Using a different block size or imposing boundary conditions, e.g. limiting the gap width, would lead to different fittest structures.

We now consider the near-field intensity enhancement maps of both reference and matrix antenna in Fig.~\ref{fig:comparison}(c).  The small displacement of the rod-like structures increases the near-field intensity enhancement by a small factor because of the proximity of the corners of the rod-like structures to the point of optimization. However, this alone by far cannot explain the observed increase of the near-field intensity enhancement. The achievable enhancement by displacing the reference antenna in a similar way amounts to a factor of 1.1.

The result of a \emph{toggle plot analysis} is displayed in Fig.~\ref{fig:comparison}(d). It indicates that changing individual blocks does not yield considerable additional near-field intensity enhancement, but rather a severe reduction. We therefore believe that the structure's fitness is close to a (local) maximum in the configuration space. The by far strongest reduction of fitness occurs when toggling gold blocks near the center. This indicates that the compact structure in the proximity of the gap is dominating the field enhancement and is most critical for achieving the observed performance. Assuming that it is excited at an Eigenmode, this also explains the appearance of a single narrow Lorentzian resonance. As apparent from the toggle plot, the random structures far away from the center do hardly influence the field enhancement in the gap. Nevertheless, it is possible that collective effects of the peripheral blocks do influence the fitness of the structure to some extend which due to the inherent complexity will not be further discussed in detail. 

\begin{figure*}[htp]%
\centering
\includegraphics[width=\textwidth]{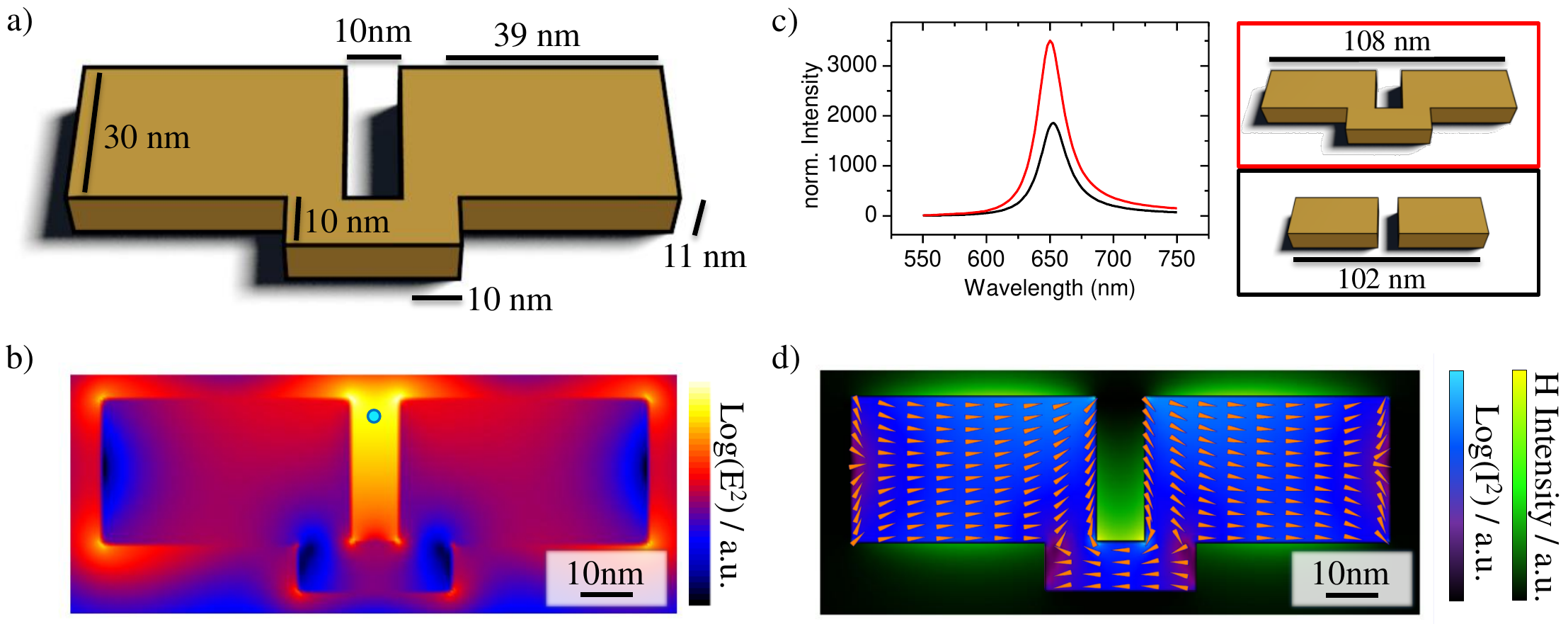}%
\caption{\label{fig:split-ringAntenna} (color online) (a) Geometry and dimensions of the examined split-ring antenna. (b) Near field intensity of the same structure, showing the very concentrated fields in the open end of the gap. The blue point denotes the the point of measurement for the spectrum figured in (c). The $Q \approx\;$25 for the SRA (red) is comparable to the reference antenna (black).  (d) is an overlay of the current intensities and direction inside the material, and the magnetic near fields outside the structure.}%
\end{figure*}

Of particular interest is the single gold block below the gap. It provides a current path via the cube edges between the two rod-like structures that form the gap as can be seen by taking a closer look at the currents in the central part of the matrix antenna in Fig.~\ref{fig:comparison}(e). Surprisingly, we find that removing this block severely lowers the fitness of the resulting matrix antenna instead of increasing it. Closer inspection of Fig.~\ref{fig:comparison}(e) reveals two particular current paths, one located in the rod-like structures corresponding to a bonding linear dipolar nano antenna mode, but also a second one, which flows from one upper gap edge through the connecting gold block to the other upper gap edge, corresponding to a fundamental split ring mode.

In order to better understand the effects that lead to the increased near-field intensity enhancement, in the following we study a reduced model system, i.e.~a mixture of a split-ring and a linear two-wire antenna called \emph{split-ring antenna} (SRA) that retains the important features of the fittest matrix antenna but can be described by a small number of freely tunable parameters. Its geometry is depicted in Fig.~\ref{fig:split-ringAntenna}(a). The structure can be interpreted either as linear two-wire antenna with an asymmetric short circuit, a split-ring resonator with attached wires or a long single nano wire that is deformed in a particular way.

The near-field intensity enhancement of the SRA in the center plane is depicted in Fig.~\ref{fig:split-ringAntenna}(b), showing a strong field concentration towards the open side of the SRA gap as it was already observed in the best matrix antenna. Fig.~\ref{fig:split-ringAntenna}(c) compares the spectra of the resonances of a SRA with a non short-circuited dipolar antenna of identical arm cross section, gap size and resonance frequency. Both spectra were obtained at the point of highest near field enhancement along the $y$-axis. Also in the present model system the split-ring antenna surpasses the classical dipole antenna design in terms of maximum near-field intensity enhancement by a factor of 2.

Also in the model system the current pattern can be decomposed into a fundamental (n=1) split-ring mode \cite{Rockstuhl2006} and a dipolar current in each antenna arm which is running 180$^\circ$ out-of-phase to the current in the short circuit (Fig.~\ref{fig:split-ringAntenna} (d)), adding to the charge accumulation in the upper part of the gap and thus increasing the near-field intensity enhancement. Since the resonance is in the optical regime, the SRA is a way to circumvent the limitation of pure split-ring resonances to wavelengths above 900 nm due to the kinetic inductance \cite{Zhou2005}. The SRA represents a magnetic dipole in the visible, showing magnetic fields (Fig.~\ref{fig:split-ringAntenna} (d)), which are only by a factor of 2.5 weaker than those of the isolated split-ring resonator at its resonance wavelength of 908~nm.

\begin{figure}[htp]%
\centering
\includegraphics[width=\columnwidth]{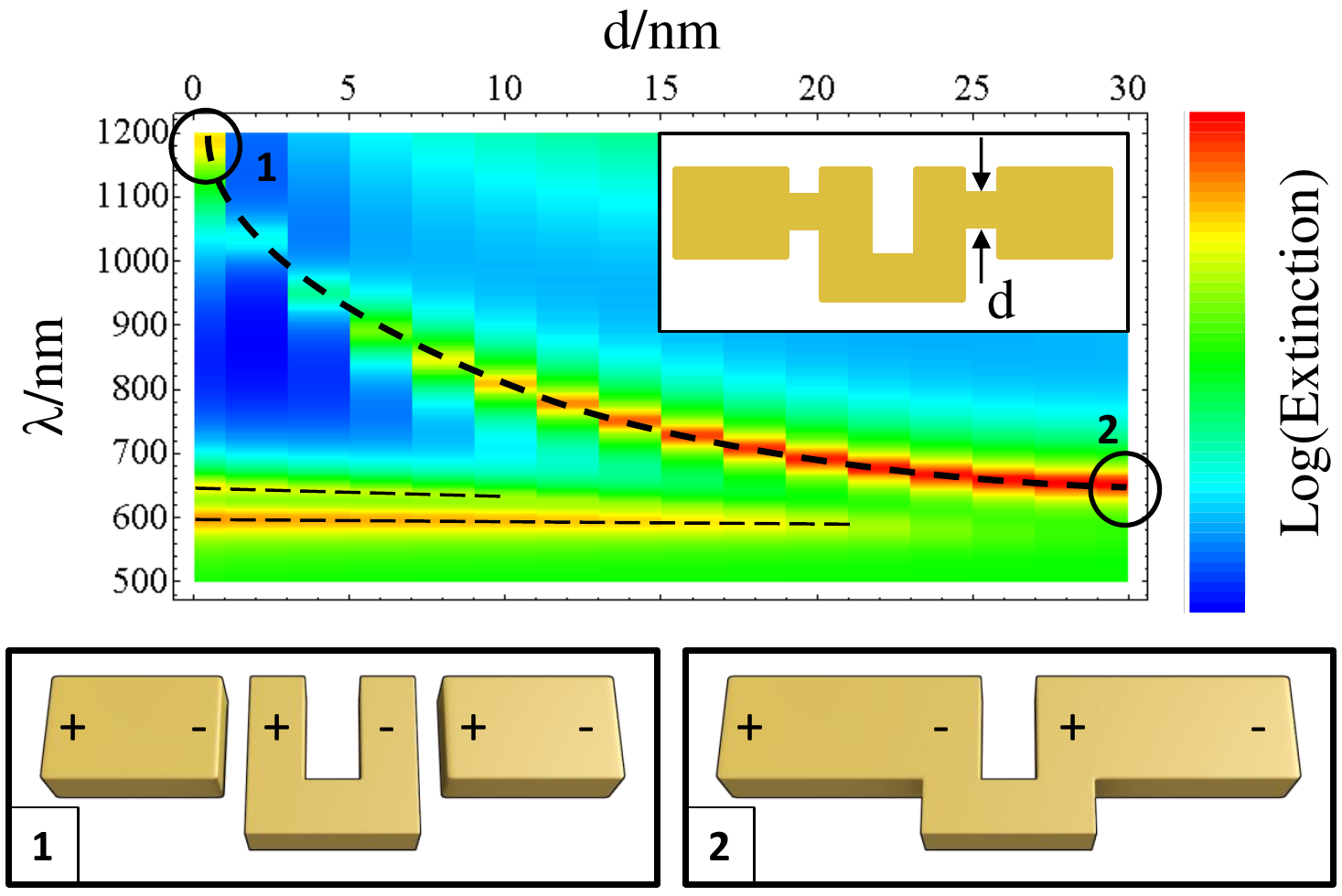}%
\caption{\label{fig:SR_Shift} Wavelength shift of the extinction cross section resonance for the combined two-wire antenna and the $n=1$ split ring resonance, as a conductive link between two antenna rods and a split ring (see inset) increases in thickness (marked thick dashed). Also visible are the bonding and antibonding $n=3$-modes, which also shift slightly into the blue and get very weak (marked thin dashed). The sketches at the bottom show the geometries of the fully disconnected and fully connected structures together with the position and sign of their mode's charge density maxima.}%
\end{figure}

To confirm the shift of the fundamental split-ring mode from the infrared into the visible spectral range, we place two gold bars (35x30x11 nm$^3$) separated by a 5~nm non conductive gap from the plain split-rings ends.  The extinction cross section is examined while connecting the gold bars with the split-ring via successively thicker gold bridges at the center of the gap (see Fig.~\ref{fig:SR_Shift} and inset). In the unconnected geometry the fundamental split-ring resonance is shifted into the red to about 1200 nm due to the capacitive coupling across the gap. As the connection grows thicker (increasing $d$), the fundamental split-ring mode is shifted by more than 500~nm from the infrared into the visible. The shift of the resonance is of similar origin as the emergence of a charge transfer mode for a dipolar antenna with a conductive bridge \cite{Schnell2009}. However, here the fundamental split-ring mode is not disappearing but its phase is inverted (compare the charge distributions sketched in the lower panel of Fig.~\ref{fig:SR_Shift}). For very thin conductive bridges both modes exist simultaneously (see also \cite{Schnell2009}) and cancel out each other, leading to a dip in the extinction cross section. 

\paragraph{Conclusion}
We show that by using the method of evolutionary optimization in a large parameter space, high-fitness plasmonic antennas can be found within a reasonable amount of time. The method can be adapted to a large variety of fitness parameters in order to optimize plasmonic structures for various purposes. Besides directly yielding optimized structures, careful analysis of the working principles of the resulting geometries may provide new design strategies for high-performance plasmonic nano structures.

In the present example we obtain an increase of the near-field intensity by nearly a factor of two caused by the intriguing cooperation of a fundamental split-ring mode and dipole antenna resonance. We have used the latter principle to devise a novel antenna design which additionally exhibits very large magnetic fields at optical frequencies. In particular, the fundamental split-ring resonance is shifted into the visible spectral range because of the formation of a charge-transfer-like hybrid resonance with two rods of a dipolar antenna. The method can be further adapted to include geometrical constraints imposed by micro fabrication and therefore can lead to structures that can directly be implemented in practical applications.

\paragraph{Acknowledgements}
Markus Kiunke wants to thank Christoph Br\"uning, Thorsten Feichtner and Oleg Selig want to thank Jord C. Prangsma and Paolo Biagioni for fruitful discussions. Financial support by the DFG is gratefully acknowledged (HE5618/1-1).


\begin{thebibliography}{37}%
\makeatletter
\providecommand \@ifxundefined [1]{%
 \@ifx{#1\undefined}
}%
\providecommand \@ifnum [1]{%
 \ifnum #1\expandafter \@firstoftwo
 \else \expandafter \@secondoftwo
 \fi
}%
\providecommand \@ifx [1]{%
 \ifx #1\expandafter \@firstoftwo
 \else \expandafter \@secondoftwo
 \fi
}%
\providecommand \natexlab [1]{#1}%
\providecommand \enquote  [1]{``#1''}%
\providecommand \bibnamefont  [1]{#1}%
\providecommand \bibfnamefont [1]{#1}%
\providecommand \citenamefont [1]{#1}%
\providecommand \href@noop [0]{\@secondoftwo}%
\providecommand \href [0]{\begingroup \@sanitize@url \@href}%
\providecommand \@href[1]{\@@startlink{#1}\@@href}%
\providecommand \@@href[1]{\endgroup#1\@@endlink}%
\providecommand \@sanitize@url [0]{\catcode `\\12\catcode `\$12\catcode
  `\&12\catcode `\#12\catcode `\^12\catcode `\_12\catcode `\%12\relax}%
\providecommand \@@startlink[1]{}%
\providecommand \@@endlink[0]{}%
\providecommand \url  [0]{\begingroup\@sanitize@url \@url }%
\providecommand \@url [1]{\endgroup\@href {#1}{\urlprefix }}%
\providecommand \urlprefix  [0]{URL }%
\providecommand \Eprint [0]{\href }%
\providecommand \doibase [0]{http://dx.doi.org/}%
\providecommand \selectlanguage [0]{\@gobble}%
\providecommand \bibinfo  [0]{\@secondoftwo}%
\providecommand \bibfield  [0]{\@secondoftwo}%
\providecommand \translation [1]{[#1]}%
\providecommand \BibitemOpen [0]{}%
\providecommand \bibitemStop [0]{}%
\providecommand \bibitemNoStop [0]{.\EOS\space}%
\providecommand \EOS [0]{\spacefactor3000\relax}%
\providecommand \BibitemShut  [1]{\csname bibitem#1\endcsname}%
\let\auto@bib@innerbib\@empty
\bibitem [{\citenamefont {Bharadwaj}\ \emph {et~al.}(2009)\citenamefont {Deutsch},\ and\ \citenamefont {Novotny}}]{Bharadwaj2009}%
  \BibitemOpen
  \bibfield  {author} {\bibinfo {author} {\bibfnamefont {P.}~\bibnamefont
  {Bharadwaj}}, \bibinfo {author} {\bibfnamefont {B.}~\bibnamefont {Deutsch}}, \ and\ 
  \bibinfo {author} {\bibfnamefont {L.}~\bibnamefont {Novotny}}, }\href@noop {} {\bibfield  {journal} {\bibinfo  {journal} {Adv. Opt. Phot.}\ }\textbf {\bibinfo {volume}
  {1}},\ \bibinfo {pages} {438--483} (\bibinfo {year} {2009})}\BibitemShut
  {NoStop}%
\bibitem [{\citenamefont {Biagioni}\ \emph {et~al.}(2012)\citenamefont
  {Biagioni}, \citenamefont {Huang},\ and\ \citenamefont
  {Hecht}}]{Biagioni2012}%
  \BibitemOpen
  \bibfield  {author} {\bibinfo {author} {\bibfnamefont {P.}~\bibnamefont
  {Biagioni}}, \bibinfo {author} {\bibfnamefont {J.-S.}\ \bibnamefont {Huang}},
  \ and\ \bibinfo {author} {\bibfnamefont {B.}~\bibnamefont {Hecht}},\
  }\href@noop {} {\bibfield  {journal} {\bibinfo  {journal} {Rep. Prog. Phys.}\
  }\textbf {\bibinfo {volume} {75}},\ \bibinfo {pages} {024402} (\bibinfo
  {year} {2012})}\BibitemShut {NoStop}%
\bibitem [{\citenamefont {Atwater}\ and\ \citenamefont
  {Polman}(2010)}]{Atwater2010}%
  \BibitemOpen
  \bibfield  {author} {\bibinfo {author} {\bibfnamefont {H.~A.}\ \bibnamefont
  {Atwater}}\ and\ \bibinfo {author} {\bibfnamefont {A.}~\bibnamefont
  {Polman}},\ }\href@noop {} {\bibfield  {journal} {\bibinfo  {journal} {Nat.
  Mater.}\ }\textbf {\bibinfo {volume} {9}},\ \bibinfo {pages} {205} (\bibinfo
  {year} {2010})}\BibitemShut {NoStop}%
\bibitem [{\citenamefont {Liu}\ \emph {et~al.}(2011)\citenamefont {Liu},
  \citenamefont {Hou}, \citenamefont {Pavaskar}, \citenamefont {Aykol},\ and\
  \citenamefont {Cronin}}]{Liu2011a}%
  \BibitemOpen
  \bibfield  {author} {\bibinfo {author} {\bibfnamefont {Z.}~\bibnamefont
  {Liu}}, \bibinfo {author} {\bibfnamefont {W.}~\bibnamefont {Hou}}, \bibinfo
  {author} {\bibfnamefont {P.}~\bibnamefont {Pavaskar}}, \bibinfo {author}
  {\bibfnamefont {M.}~\bibnamefont {Aykol}}, \ and\ \bibinfo {author}
  {\bibfnamefont {S.~B.}\ \bibnamefont {Cronin}},\ }\href {\doibase
  10.1021/nl104005n} {\bibfield  {journal} {\bibinfo  {journal} {Nano Letters}\
  }\textbf {\bibinfo {volume} {11}},\ \bibinfo {pages} {1111} (\bibinfo {year}
  {2011})},\ \Eprint
  {http://arxiv.org/abs/http://pubs.acs.org/doi/pdf/10.1021/nl104005n}
  {http://pubs.acs.org/doi/pdf/10.1021/nl104005n} \BibitemShut {NoStop}%
\bibitem [{\citenamefont {Becker}\ \emph {et~al.}(2010)\citenamefont {Becker},
  \citenamefont {Tr\"{u}gler}, \citenamefont {Jakab}, \citenamefont
  {Hohenester},\ and\ \citenamefont {S\"{o}nnichsen}}]{Becker2010}%
  \BibitemOpen
  \bibfield  {author} {\bibinfo {author} {\bibfnamefont {J.}~\bibnamefont
  {Becker}}, \bibinfo {author} {\bibfnamefont {A.}~\bibnamefont {Tr\"{u}gler}},
  \bibinfo {author} {\bibfnamefont {A.}~\bibnamefont {Jakab}}, \bibinfo
  {author} {\bibfnamefont {U.}~\bibnamefont {Hohenester}}, \ and\ \bibinfo
  {author} {\bibfnamefont {C.}~\bibnamefont {S\"{o}nnichsen}},\ }\href
  {\doibase 10.1007/s11468-010-9130-2} {\bibfield  {journal} {\bibinfo
  {journal} {Plasmonics}\ }\textbf {\bibinfo {volume} {5}},\ \bibinfo {pages}
  {161} (\bibinfo {year} {2010})}\BibitemShut {NoStop}%
\bibitem [{\citenamefont {Juan}\ \emph {et~al.}(2009)\citenamefont {Juan},
  \citenamefont {Gordon}, \citenamefont {Pang}, \citenamefont {Eftekhari},\
  and\ \citenamefont {Quidant}}]{Juan2009}%
  \BibitemOpen
  \bibfield  {author} {\bibinfo {author} {\bibfnamefont {M.~L.}\ \bibnamefont
  {Juan}}, \bibinfo {author} {\bibfnamefont {R.}~\bibnamefont {Gordon}},
  \bibinfo {author} {\bibfnamefont {Y.}~\bibnamefont {Pang}}, \bibinfo {author}
  {\bibfnamefont {F.}~\bibnamefont {Eftekhari}}, \ and\ \bibinfo {author}
  {\bibfnamefont {R.}~\bibnamefont {Quidant}},\ }\href {\doibase
  10.1038/nphys1422} {\bibfield  {journal} {\bibinfo  {journal} {Nat. Phys.}\
  }\textbf {\bibinfo {volume} {5}},\ \bibinfo {pages} {915} (\bibinfo {year}
  {2009})}\BibitemShut {NoStop}%
\bibitem [{\citenamefont {Zhang}\ \emph {et~al.}(2010)\citenamefont {Zhang},
  \citenamefont {Huang}, \citenamefont {Santschi},\ and\ \citenamefont
  {Martin}}]{Zhang2010}%
  \BibitemOpen
  \bibfield  {author} {\bibinfo {author} {\bibfnamefont {W.}~\bibnamefont
  {Zhang}}, \bibinfo {author} {\bibfnamefont {L.}~\bibnamefont {Huang}},
  \bibinfo {author} {\bibfnamefont {C.}~\bibnamefont {Santschi}}, \ and\
  \bibinfo {author} {\bibfnamefont {O.~J.~F.}\ \bibnamefont {Martin}},\ }\href
  {http://dx.doi.org/10.1021/nl904168f} {\bibfield  {journal} {\bibinfo
  {journal} {Nano Letters}\ }\textbf {\bibinfo {volume} {10}},\ \bibinfo
  {pages} {1006} (\bibinfo {year} {2010})}\BibitemShut {NoStop}%
\bibitem [{\citenamefont {Kinkhabwala}\ \emph {et~al.}(2009)\citenamefont
  {Kinkhabwala}, \citenamefont {Yu}, \citenamefont {Fan}, \citenamefont
  {Avlasevich}, \citenamefont {Muellen},\ and\ \citenamefont
  {Moerner}}]{Kinkhabwala2009}%
  \BibitemOpen
  \bibfield  {author} {\bibinfo {author} {\bibfnamefont {A.}~\bibnamefont
  {Kinkhabwala}}, \bibinfo {author} {\bibfnamefont {Z.}~\bibnamefont {Yu}},
  \bibinfo {author} {\bibfnamefont {S.}~\bibnamefont {Fan}}, \bibinfo {author}
  {\bibfnamefont {Y.}~\bibnamefont {Avlasevich}}, \bibinfo {author}
  {\bibfnamefont {K.}~\bibnamefont {Muellen}}, \ and\ \bibinfo {author}
  {\bibfnamefont {W.~E.}\ \bibnamefont {Moerner}},\ }\href {\doibase
  10.1038/NPHOTON.2009.187} {\bibfield  {journal} {\bibinfo  {journal} {Nat.
  Photon.}\ }\textbf {\bibinfo {volume} {3}},\ \bibinfo {pages} {654} (\bibinfo
  {year} {2009})}\BibitemShut {NoStop}%
\bibitem [{\citenamefont {Akimov}\ \emph {et~al.}(2007)\citenamefont {Akimov},
  \citenamefont {Mukherjee}, \citenamefont {Yu}, \citenamefont {Chang},
  \citenamefont {Zibrov}, \citenamefont {Hemmer}, \citenamefont {Park},\ and\
  \citenamefont {Lukin}}]{Akimov2007}%
  \BibitemOpen
  \bibfield  {author} {\bibinfo {author} {\bibfnamefont {a.~V.}\ \bibnamefont
  {Akimov}}, \bibinfo {author} {\bibfnamefont {a.}~\bibnamefont {Mukherjee}},
  \bibinfo {author} {\bibfnamefont {C.~L.}\ \bibnamefont {Yu}}, \bibinfo
  {author} {\bibfnamefont {D.~E.}\ \bibnamefont {Chang}}, \bibinfo {author}
  {\bibfnamefont {a.~S.}\ \bibnamefont {Zibrov}}, \bibinfo {author}
  {\bibfnamefont {P.~R.}\ \bibnamefont {Hemmer}}, \bibinfo {author}
  {\bibfnamefont {H.}~\bibnamefont {Park}}, \ and\ \bibinfo {author}
  {\bibfnamefont {M.~D.}\ \bibnamefont {Lukin}},\ }\href {\doibase
  10.1038/nature06230} {\bibfield  {journal} {\bibinfo  {journal} {Nature}\
  }\textbf {\bibinfo {volume} {450}},\ \bibinfo {pages} {402} (\bibinfo {year}
  {2007})}\BibitemShut {NoStop}%
\bibitem [{\citenamefont {Kolesov}\ \emph {et~al.}(2009)\citenamefont
  {Kolesov}, \citenamefont {Grotz}, \citenamefont {Balasubramanian},
  \citenamefont {Stohr}, \citenamefont {Nicolet}, \citenamefont {Hemmer},
  \citenamefont {Jelezko},\ and\ \citenamefont {Wrachtrup}}]{Kolesov2009}%
  \BibitemOpen
  \bibfield  {author} {\bibinfo {author} {\bibfnamefont {R.}~\bibnamefont
  {Kolesov}}, \bibinfo {author} {\bibfnamefont {B.}~\bibnamefont {Grotz}},
  \bibinfo {author} {\bibfnamefont {G.}~\bibnamefont {Balasubramanian}},
  \bibinfo {author} {\bibfnamefont {R.~J.}\ \bibnamefont {Stohr}}, \bibinfo
  {author} {\bibfnamefont {A.~A.~L.}\ \bibnamefont {Nicolet}}, \bibinfo
  {author} {\bibfnamefont {P.~R.}\ \bibnamefont {Hemmer}}, \bibinfo {author}
  {\bibfnamefont {F.}~\bibnamefont {Jelezko}}, \ and\ \bibinfo {author}
  {\bibfnamefont {J.}~\bibnamefont {Wrachtrup}},\ }\href
  {http://dx.doi.org/10.1038/nphys1278} {\bibfield  {journal} {\bibinfo
  {journal} {Nat. Phys.}\ }\textbf {\bibinfo {volume} {5}},\ \bibinfo {pages}
  {470} (\bibinfo {year} {2009})}\BibitemShut {NoStop}%
\bibitem [{\citenamefont {Huang}\ \emph {et~al.}(2009)\citenamefont {Huang},
  \citenamefont {Feichtner}, \citenamefont {Biagioni},\ and\ \citenamefont
  {Hecht}}]{Huang2009}%
  \BibitemOpen
  \bibfield  {author} {\bibinfo {author} {\bibfnamefont {J.-S.}\ \bibnamefont
  {Huang}}, \bibinfo {author} {\bibfnamefont {T.}~\bibnamefont {Feichtner}},
  \bibinfo {author} {\bibfnamefont {P.}~\bibnamefont {Biagioni}}, \ and\
  \bibinfo {author} {\bibfnamefont {B.}~\bibnamefont {Hecht}},\ }\href
  {\doibase 10.1021/nl803902t} {\bibfield  {journal} {\bibinfo  {journal} {Nano
  Letters}\ }\textbf {\bibinfo {volume} {9}},\ \bibinfo {pages} {1897}
  (\bibinfo {year} {2009})}\BibitemShut {NoStop}%
\bibitem [{\citenamefont {Jacob}\ and\ \citenamefont
  {Shalaev}(2011)}]{Jacob2011}%
  \BibitemOpen
  \bibfield  {author} {\bibinfo {author} {\bibfnamefont {Z.}~\bibnamefont
  {Jacob}}\ and\ \bibinfo {author} {\bibfnamefont {V.~M.}\ \bibnamefont
  {Shalaev}},\ }\href {\doibase 10.1126/science.1211736} {\bibfield  {journal}
  {\bibinfo  {journal} {Science}\ }\textbf {\bibinfo {volume} {334}},\ \bibinfo
  {pages} {463} (\bibinfo {year} {2011})}\BibitemShut {NoStop}%
\bibitem [{\citenamefont {Balanis}(1992)}]{Balanis1992}%
  \BibitemOpen
  \bibfield  {author} {\bibinfo {author} {\bibfnamefont {C.~A.}\ \bibnamefont
  {Balanis}},\ }\href@noop {} {\bibfield  {journal} {\bibinfo  {journal} {Proc.
  IEEE}\ }\textbf {\bibinfo {volume} {80}},\ \bibinfo {pages} {7} (\bibinfo
  {year} {1992})}\BibitemShut {NoStop}%
\bibitem [{\citenamefont {Dorfmueller}\ \emph {et~al.}(2010)\citenamefont
  {Dorfmueller}, \citenamefont {Vogelgesang}, \citenamefont {Khunsin},
  \citenamefont {Rockstuhl}, \citenamefont {Etrich},\ and\ \citenamefont
  {Kern}}]{Dorfmueller2010}%
  \BibitemOpen
  \bibfield  {author} {\bibinfo {author} {\bibfnamefont {J.}~\bibnamefont
  {Dorfmueller}}, \bibinfo {author} {\bibfnamefont {R.}~\bibnamefont
  {Vogelgesang}}, \bibinfo {author} {\bibfnamefont {W.}~\bibnamefont
  {Khunsin}}, \bibinfo {author} {\bibfnamefont {C.}~\bibnamefont {Rockstuhl}},
  \bibinfo {author} {\bibfnamefont {C.}~\bibnamefont {Etrich}}, \ and\ \bibinfo
  {author} {\bibfnamefont {K.}~\bibnamefont {Kern}},\ }\href@noop {} {\bibfield
   {journal} {\bibinfo  {journal} {Nano Letters}\ }\textbf {\bibinfo {volume}
  {10}},\ \bibinfo {pages} {3596} (\bibinfo {year} {2010})}\BibitemShut
  {NoStop}%
\bibitem [{\citenamefont {Novotny}(2007)}]{Novotny2007}%
  \BibitemOpen
  \bibfield  {author} {\bibinfo {author} {\bibfnamefont {L.}~\bibnamefont
  {Novotny}},\ }\href {\doibase 10.1103/PhysRevLett.98.266802} {\bibfield
  {journal} {\bibinfo  {journal} {Phys. Rev. Lett.}\ }\textbf {\bibinfo
  {volume} {98}},\ \bibinfo {eid} {266802} (\bibinfo {year}
  {2007})}\BibitemShut {NoStop}%
\bibitem [{\citenamefont {Zhou}\ \emph {et~al.}(2005)\citenamefont {Zhou},
  \citenamefont {Koschny}, \citenamefont {Kafesaki}, \citenamefont {Economou},
  \citenamefont {Pendry},\ and\ \citenamefont {Soukoulis}}]{Zhou2005}%
  \BibitemOpen
  \bibfield  {author} {\bibinfo {author} {\bibfnamefont {J.}~\bibnamefont
  {Zhou}}, \bibinfo {author} {\bibfnamefont {T.}~\bibnamefont {Koschny}},
  \bibinfo {author} {\bibfnamefont {M.}~\bibnamefont {Kafesaki}}, \bibinfo
  {author} {\bibfnamefont {E.~N.}\ \bibnamefont {Economou}}, \bibinfo {author}
  {\bibfnamefont {J.~B.}\ \bibnamefont {Pendry}}, \ and\ \bibinfo {author}
  {\bibfnamefont {C.~M.}\ \bibnamefont {Soukoulis}},\ }\href {\doibase
  10.1103/PhysRevLett.95.223902} {\bibfield  {journal} {\bibinfo  {journal}
  {Phys. Rev. Lett.}\ }\textbf {\bibinfo {volume} {95}},\ \bibinfo {pages}
  {223902} (\bibinfo {year} {2005})}\BibitemShut {NoStop}%
\bibitem [{\citenamefont {M\"uhlschlegel}\ \emph {et~al.}(2005)\citenamefont
  {M\"uhlschlegel}, \citenamefont {Eisler}, \citenamefont {Martin},
  \citenamefont {Hecht},\ and\ \citenamefont {Pohl}}]{Muehlschlegel2005}%
  \BibitemOpen
  \bibfield  {author} {\bibinfo {author} {\bibfnamefont {P.}~\bibnamefont
  {M\"uhlschlegel}}, \bibinfo {author} {\bibfnamefont {H.-J.}\ \bibnamefont
  {Eisler}}, \bibinfo {author} {\bibfnamefont {O.~J.~F.}\ \bibnamefont
  {Martin}}, \bibinfo {author} {\bibfnamefont {B.}~\bibnamefont {Hecht}}, \
  and\ \bibinfo {author} {\bibfnamefont {D.~W.}\ \bibnamefont {Pohl}},\ }\href
  {\doibase 10.1126/science.1111886} {\bibfield  {journal} {\bibinfo  {journal}
  {Science}\ }\textbf {\bibinfo {volume} {308}},\ \bibinfo {pages} {1607}
  (\bibinfo {year} {2005})}\BibitemShut {NoStop}%
\bibitem [{\citenamefont {Schuck}\ \emph {et~al.}(2005)\citenamefont {Schuck},
  \citenamefont {Fromm}, \citenamefont {Sundaramurthy}, \citenamefont {Kino},\
  and\ \citenamefont {Moerner}}]{Schuck2005}%
  \BibitemOpen
  \bibfield  {author} {\bibinfo {author} {\bibfnamefont {P.}~\bibnamefont
  {Schuck}}, \bibinfo {author} {\bibfnamefont {D.}~\bibnamefont {Fromm}},
  \bibinfo {author} {\bibfnamefont {A.}~\bibnamefont {Sundaramurthy}}, \bibinfo
  {author} {\bibfnamefont {G.}~\bibnamefont {Kino}}, \ and\ \bibinfo {author}
  {\bibfnamefont {W.}~\bibnamefont {Moerner}},\ }\href@noop {} {\bibfield
  {journal} {\bibinfo  {journal} {Phys. Rev. Lett.}\ }\textbf {\bibinfo
  {volume} {94}},\ \bibinfo {pages} {017402} (\bibinfo {year}
  {2005})}\BibitemShut {NoStop}%
\bibitem [{\citenamefont {Farahani}\ \emph {et~al.}(2005)\citenamefont
  {Farahani}, \citenamefont {Pohl}, \citenamefont {Eisler},\ and\ \citenamefont
  {Hecht}}]{Farahani2005}%
  \BibitemOpen
  \bibfield  {author} {\bibinfo {author} {\bibfnamefont {J.~N.}\ \bibnamefont
  {Farahani}}, \bibinfo {author} {\bibfnamefont {D.~W.}\ \bibnamefont {Pohl}},
  \bibinfo {author} {\bibfnamefont {H.-J.}\ \bibnamefont {Eisler}}, \ and\
  \bibinfo {author} {\bibfnamefont {B.}~\bibnamefont {Hecht}},\ }\href
  {\doibase 10.1103/PhysRevLett.95.017402} {\bibfield  {journal} {\bibinfo
  {journal} {Phys. Rev. Lett.}\ }\textbf {\bibinfo {volume} {95}},\ \bibinfo
  {pages} {017402} (\bibinfo {year} {2005})}\BibitemShut {NoStop}%
\bibitem [{\citenamefont {Taminiau}\ \emph {et~al.}(2008)\citenamefont
  {Taminiau}, \citenamefont {Stefani},\ and\ \citenamefont
  {Van~Hulst}}]{Taminiau2008}%
  \BibitemOpen
  \bibfield  {author} {\bibinfo {author} {\bibfnamefont {T.~H.}\ \bibnamefont
  {Taminiau}}, \bibinfo {author} {\bibfnamefont {F.~D.}\ \bibnamefont
  {Stefani}}, \ and\ \bibinfo {author} {\bibfnamefont {N.~F.}\ \bibnamefont
  {Van~Hulst}},\ }\href@noop {} {\bibfield  {journal} {\bibinfo  {journal}
  {Opt. Expr.}\ }\textbf {\bibinfo {volume} {16}},\ \bibinfo {pages} {10858}
  (\bibinfo {year} {2008})}\BibitemShut {NoStop}%
\bibitem [{\citenamefont {Curto}\ \emph {et~al.}(2010)\citenamefont {Curto},
  \citenamefont {Volpe}, \citenamefont {Taminiau}, \citenamefont {Kreuzer},
  \citenamefont {Quidant},\ and\ \citenamefont {van Hulst}}]{Curto2010}%
  \BibitemOpen
  \bibfield  {author} {\bibinfo {author} {\bibfnamefont {A.~G.}\ \bibnamefont
  {Curto}}, \bibinfo {author} {\bibfnamefont {G.}~\bibnamefont {Volpe}},
  \bibinfo {author} {\bibfnamefont {T.~H.}\ \bibnamefont {Taminiau}}, \bibinfo
  {author} {\bibfnamefont {M.~P.}\ \bibnamefont {Kreuzer}}, \bibinfo {author}
  {\bibfnamefont {R.}~\bibnamefont {Quidant}}, \ and\ \bibinfo {author}
  {\bibfnamefont {N.~F.}\ \bibnamefont {van Hulst}},\ }\href {\doibase
  10.1126/science.1191922} {\bibfield  {journal} {\bibinfo  {journal}
  {Science}\ }\textbf {\bibinfo {volume} {329}},\ \bibinfo {pages} {930}
  (\bibinfo {year} {2010})}\BibitemShut {NoStop}%
\bibitem [{\citenamefont {Sivanandram}\ and\ \citenamefont
  {Deepa}(2008)}]{Sivanandram2008}%
  \BibitemOpen
  \bibfield  {author} {\bibinfo {author} {\bibfnamefont {S.}~\bibnamefont
  {Sivanandram}}\ and\ \bibinfo {author} {\bibfnamefont {S.}~\bibnamefont
  {Deepa}},\ }\href@noop {} {\emph {\bibinfo {title} {Introduction into genetic
  algorithms}}}\ (\bibinfo  {publisher} {Springer Verlag Berlin Heidelberg},\
  \bibinfo {year} {2008})\BibitemShut {NoStop}%
\bibitem [{\citenamefont {Baumert}\ \emph {et~al.}(1997)\citenamefont
  {Baumert}, \citenamefont {Brixner}, \citenamefont {Seyfried}, \citenamefont
  {Strehle},\ and\ \citenamefont {Gerber}}]{Baumert1997}%
  \BibitemOpen
  \bibfield  {author} {\bibinfo {author} {\bibfnamefont {T.}~\bibnamefont
  {Baumert}}, \bibinfo {author} {\bibfnamefont {T.}~\bibnamefont {Brixner}},
  \bibinfo {author} {\bibfnamefont {V.}~\bibnamefont {Seyfried}}, \bibinfo
  {author} {\bibfnamefont {M.}~\bibnamefont {Strehle}}, \ and\ \bibinfo
  {author} {\bibfnamefont {G.}~\bibnamefont {Gerber}},\ }\href@noop {}
  {\bibfield  {journal} {\bibinfo  {journal} {Appl. Phys. B: Lasers and
  Optics}\ }\textbf {\bibinfo {volume} {65}},\ \bibinfo {pages} {779} (\bibinfo
  {year} {1997})}\BibitemShut {NoStop}%
\bibitem [{\citenamefont {Aeschlimann}\ \emph {et~al.}(2007)\citenamefont
  {Aeschlimann}, \citenamefont {Bauer}, \citenamefont {Bayer}, \citenamefont
  {Brixner}, \citenamefont {{Garc\'{\i}a de Abajo}}, \citenamefont {Pfeiffer},
  \citenamefont {Rohmer}, \citenamefont {Spindler},\ and\ \citenamefont
  {Steeb}}]{Aeschlimann2007}%
  \BibitemOpen
  \bibfield  {author} {\bibinfo {author} {\bibfnamefont {M.}~\bibnamefont
  {Aeschlimann}}, \bibinfo {author} {\bibfnamefont {M.}~\bibnamefont {Bauer}},
  \bibinfo {author} {\bibfnamefont {D.}~\bibnamefont {Bayer}}, \bibinfo
  {author} {\bibfnamefont {T.}~\bibnamefont {Brixner}}, \bibinfo {author}
  {\bibfnamefont {F.~J.}\ \bibnamefont {{Garc\'{\i}a de Abajo}}}, \bibinfo
  {author} {\bibfnamefont {W.}~\bibnamefont {Pfeiffer}}, \bibinfo {author}
  {\bibfnamefont {M.}~\bibnamefont {Rohmer}}, \bibinfo {author} {\bibfnamefont
  {C.}~\bibnamefont {Spindler}}, \ and\ \bibinfo {author} {\bibfnamefont
  {F.}~\bibnamefont {Steeb}},\ }\href {\doibase 10.1038/nature05595} {\bibfield
   {journal} {\bibinfo  {journal} {Nature}\ }\textbf {\bibinfo {volume}
  {446}},\ \bibinfo {pages} {301} (\bibinfo {year} {2007})}\BibitemShut
  {NoStop}%
\bibitem [{\citenamefont {Aeschlimann}\ \emph {et~al.}(2010)\citenamefont
  {Aeschlimann}, \citenamefont {Bauer}, \citenamefont {Bayer}, \citenamefont
  {Brixner}, \citenamefont {Cunovic}, \citenamefont {Dimler}, \citenamefont
  {Fischer}, \citenamefont {Pfeiffer}, \citenamefont {Rohmer}, \citenamefont
  {Schneider}, \citenamefont {Steeb}, \citenamefont {Str\"uber},\ and\
  \citenamefont {Voronine}}]{Aeschlimann2010}%
  \BibitemOpen
  \bibfield  {author} {\bibinfo {author} {\bibfnamefont {M.}~\bibnamefont
  {Aeschlimann}}, \bibinfo {author} {\bibfnamefont {M.}~\bibnamefont {Bauer}},
  \bibinfo {author} {\bibfnamefont {D.}~\bibnamefont {Bayer}}, \bibinfo
  {author} {\bibfnamefont {T.}~\bibnamefont {Brixner}}, \bibinfo {author}
  {\bibfnamefont {S.}~\bibnamefont {Cunovic}}, \bibinfo {author} {\bibfnamefont
  {F.}~\bibnamefont {Dimler}}, \bibinfo {author} {\bibfnamefont
  {A.}~\bibnamefont {Fischer}}, \bibinfo {author} {\bibfnamefont
  {W.}~\bibnamefont {Pfeiffer}}, \bibinfo {author} {\bibfnamefont
  {M.}~\bibnamefont {Rohmer}}, \bibinfo {author} {\bibfnamefont
  {C.}~\bibnamefont {Schneider}}, \bibinfo {author} {\bibfnamefont
  {F.}~\bibnamefont {Steeb}}, \bibinfo {author} {\bibfnamefont
  {C.}~\bibnamefont {Str\"uber}}, \ and\ \bibinfo {author} {\bibfnamefont
  {D.~V.}\ \bibnamefont {Voronine}},\ }\href {\doibase 10.1073/pnas.0913556107}
  {\bibfield  {journal} {\bibinfo  {journal} {Proc. Natl. Acad. of Sci. USA}\
  }\textbf {\bibinfo {volume} {107}},\ \bibinfo {pages} {5329} (\bibinfo {year}
  {2010})}\BibitemShut {NoStop}%
\bibitem [{\citenamefont {Huang}\ \emph {et~al.}(2007)\citenamefont {Huang},
  \citenamefont {Hoorfar},\ and\ \citenamefont {Lakhani}}]{Huang2007}%
  \BibitemOpen
  \bibfield  {author} {\bibinfo {author} {\bibfnamefont {H.}~\bibnamefont
  {Huang}}, \bibinfo {author} {\bibfnamefont {A.}~\bibnamefont {Hoorfar}}, \
  and\ \bibinfo {author} {\bibfnamefont {S.}~\bibnamefont {Lakhani}},\
  }\href@noop {} {\bibfield  {journal} {\bibinfo  {journal} {Antennas and
  Propagation Society International Symposium, 2007 IEEE}\ }\textbf {\bibinfo
  {volume} {1}},\ \bibinfo {pages} {1609 } (\bibinfo {year}
  {2007})}\BibitemShut {NoStop}%
\bibitem [{\citenamefont {Pantoja}\ \emph {et~al.}(2007)\citenamefont
  {Pantoja}, \citenamefont {Bretones}, \citenamefont {Member}, \citenamefont
  {Martin},\ and\ \citenamefont {Member}}]{Pantoja2007}%
  \BibitemOpen
  \bibfield  {author} {\bibinfo {author} {\bibfnamefont {M.~F.}\ \bibnamefont
  {Pantoja}}, \bibinfo {author} {\bibfnamefont {A.~R.}\ \bibnamefont
  {Bretones}}, \bibinfo {author} {\bibfnamefont {S.}~\bibnamefont {Member}},
  \bibinfo {author} {\bibfnamefont {R.~G.}\ \bibnamefont {Martin}}, \ and\
  \bibinfo {author} {\bibfnamefont {S.}~\bibnamefont {Member}},\ }\href@noop {}
  {\bibfield  {journal} {\bibinfo  {journal} {IEEE Trans. Antennas Propag.}\
  }\textbf {\bibinfo {volume} {55}},\ \bibinfo {pages} {1111} (\bibinfo {year}
  {2007})}\BibitemShut {NoStop}%
\bibitem [{\citenamefont {Ginzburg}\ \emph {et~al.}(2011)\citenamefont
  {Ginzburg}, \citenamefont {Berkovitch}, \citenamefont {Nevet}, \citenamefont
  {Shor},\ and\ \citenamefont {Orenstein}}]{Ginzburg2011}%
  \BibitemOpen
  \bibfield  {author} {\bibinfo {author} {\bibfnamefont {P.}~\bibnamefont
  {Ginzburg}}, \bibinfo {author} {\bibfnamefont {N.}~\bibnamefont
  {Berkovitch}}, \bibinfo {author} {\bibfnamefont {A.}~\bibnamefont {Nevet}},
  \bibinfo {author} {\bibfnamefont {I.}~\bibnamefont {Shor}}, \ and\ \bibinfo
  {author} {\bibfnamefont {M.}~\bibnamefont {Orenstein}},\ }\href {\doibase
  10.1021/nl200612f} {\bibfield  {journal} {\bibinfo  {journal} {Nano Letters}\
  }\textbf {\bibinfo {volume} {11}},\ \bibinfo {pages} {2329} (\bibinfo {year}
  {2011})}\BibitemShut {NoStop}%
\bibitem [{\citenamefont {Kessentini}\ \emph {et~al.}(2011)\citenamefont
  {Kessentini}, \citenamefont {Barchiesi}, \citenamefont {Grosges},\ and\
  \citenamefont {de~la Chapelle}}]{Kessentini2011}%
  \BibitemOpen
  \bibfield  {author} {\bibinfo {author} {\bibfnamefont {S.}~\bibnamefont
  {Kessentini}}, \bibinfo {author} {\bibfnamefont {D.}~\bibnamefont
  {Barchiesi}}, \bibinfo {author} {\bibfnamefont {T.}~\bibnamefont {Grosges}},
  \ and\ \bibinfo {author} {\bibfnamefont {M.}~\bibnamefont {de~la Chapelle}},\
  }\href@noop {} {\bibfield  {journal} {\bibinfo  {journal} {Evolutionary
  Computation (CEC), 2011 IEEE Congress on}\ }\textbf {\bibinfo {volume} {1}},\
  \bibinfo {pages} {2315 } (\bibinfo {year} {2011})}\BibitemShut {NoStop}%
\bibitem [{\citenamefont {Forestiere}\ \emph {et~al.}(2010)\citenamefont
  {Forestiere}, \citenamefont {Donelli}, \citenamefont {Walsh}, \citenamefont
  {Zeni}, \citenamefont {Miano},\ and\ \citenamefont {Negro}}]{Forestiere2010}%
  \BibitemOpen
  \bibfield  {author} {\bibinfo {author} {\bibfnamefont {C.}~\bibnamefont
  {Forestiere}}, \bibinfo {author} {\bibfnamefont {M.}~\bibnamefont {Donelli}},
  \bibinfo {author} {\bibfnamefont {G.~F.}\ \bibnamefont {Walsh}}, \bibinfo
  {author} {\bibfnamefont {E.}~\bibnamefont {Zeni}}, \bibinfo {author}
  {\bibfnamefont {G.}~\bibnamefont {Miano}}, \ and\ \bibinfo {author}
  {\bibfnamefont {L.~D.}\ \bibnamefont {Negro}},\ }\href@noop {} {\bibfield
  {journal} {\bibinfo  {journal} {Opt. Lett.}\ }\textbf {\bibinfo {volume}
  {35}},\ \bibinfo {pages} {133} (\bibinfo {year} {2010})}\BibitemShut
  {NoStop}%
\bibitem [{\citenamefont {Forestiere}\ \emph {et~al.}(2012)\citenamefont
  {Forestiere}, \citenamefont {Pasquale}, \citenamefont {Capretti},
  \citenamefont {Miano}, \citenamefont {Tamburrino}, \citenamefont {Lee},
  \citenamefont {Reinhard},\ and\ \citenamefont {Dal~Negro}}]{Forestiere2012}%
  \BibitemOpen
  \bibfield  {author} {\bibinfo {author} {\bibfnamefont {C.}~\bibnamefont
  {Forestiere}}, \bibinfo {author} {\bibfnamefont {A.~J.}\ \bibnamefont
  {Pasquale}}, \bibinfo {author} {\bibfnamefont {A.}~\bibnamefont {Capretti}},
  \bibinfo {author} {\bibfnamefont {G.}~\bibnamefont {Miano}}, \bibinfo
  {author} {\bibfnamefont {A.}~\bibnamefont {Tamburrino}}, \bibinfo {author}
  {\bibfnamefont {S.~Y.}\ \bibnamefont {Lee}}, \bibinfo {author} {\bibfnamefont
  {B.~M.}\ \bibnamefont {Reinhard}}, \ and\ \bibinfo {author} {\bibfnamefont
  {L.}~\bibnamefont {Dal~Negro}},\ }\href {\doibase 10.1021/nl300140g}
  {\bibfield  {journal} {\bibinfo  {journal} {Nano Letters}\ }\textbf {\bibinfo
  {volume} {12}},\ \bibinfo {pages} {2037} (\bibinfo {year}
  {2012})}\BibitemShut {NoStop}%
\bibitem [{sup()}]{suppLink}%
  \BibitemOpen
  \href@noop {} {}\bibinfo {note} {See Supplemental Material at [URL will be
  inserted by publisher].}\BibitemShut {Stop}%
\bibitem [{\citenamefont {Taflove}\ and\ \citenamefont
  {Hagness}(2005)}]{Taflove2005}%
  \BibitemOpen
  \bibfield  {author} {\bibinfo {author} {\bibfnamefont {A.}~\bibnamefont
  {Taflove}}\ and\ \bibinfo {author} {\bibfnamefont {S.~C.}\ \bibnamefont
  {Hagness}},\ }\href@noop {} {\emph {\bibinfo {title} {Computational
  electrodynamics: the finite-difference time-domain method}}},\ \bibinfo
  {edition} {3rd}\ ed.,\ edited by\ \bibinfo {editor} {\bibfnamefont
  {A.}~\bibnamefont {Taflove}}\ (\bibinfo  {publisher} {Artech House, Inc.},\
  \bibinfo {year} {2005})\BibitemShut {NoStop}%
\bibitem [{\citenamefont {Etchegoin}\ \emph {et~al.}(2006)\citenamefont
  {Etchegoin}, \citenamefont {Ru},\ and\ \citenamefont
  {Meyer}}]{Etchegoin2006}%
  \BibitemOpen
  \bibfield  {author} {\bibinfo {author} {\bibfnamefont {P.~G.}\ \bibnamefont
  {Etchegoin}}, \bibinfo {author} {\bibfnamefont {E.~C.~L.}\ \bibnamefont
  {Ru}}, \ and\ \bibinfo {author} {\bibfnamefont {M.}~\bibnamefont {Meyer}},\
  }\href {\doibase 10.1063/1.2360270} {\bibfield  {journal} {\bibinfo
  {journal} {J. Chem. Phys}\ }\textbf {\bibinfo {volume} {125}},\ \bibinfo
  {eid} {164705} (\bibinfo {year} {2006})}\BibitemShut {NoStop}%
\bibitem [{Note1()}]{Note1}%
  \BibitemOpen
  \bibinfo {note} {All necessary code written in Lumerical script language or
  MatLab can be made available upon request.}\BibitemShut {Stop}%
\bibitem [{Note2()}]{Note2}%
  \BibitemOpen
  \bibinfo {note} {This is the brute force realization of the hill climber
  algorithm on our setup, neglecting pair- and higher-order
  correlations.}\BibitemShut {Stop}%
\bibitem [{\citenamefont {Rockstuhl}\ \emph {et~al.}(2006)\citenamefont
  {Rockstuhl}, \citenamefont {Lederer}, \citenamefont {Etrich}, \citenamefont
  {Zentgraf}, \citenamefont {Kuhl},\ and\ \citenamefont
  {Giessen}}]{Rockstuhl2006}%
  \BibitemOpen
  \bibfield  {author} {\bibinfo {author} {\bibfnamefont {C.}~\bibnamefont
  {Rockstuhl}}, \bibinfo {author} {\bibfnamefont {F.}~\bibnamefont {Lederer}},
  \bibinfo {author} {\bibfnamefont {C.}~\bibnamefont {Etrich}}, \bibinfo
  {author} {\bibfnamefont {T.}~\bibnamefont {Zentgraf}}, \bibinfo {author}
  {\bibfnamefont {J.}~\bibnamefont {Kuhl}}, \ and\ \bibinfo {author}
  {\bibfnamefont {H.}~\bibnamefont {Giessen}},\ }\href@noop {} {\bibfield
  {journal} {\bibinfo  {journal} {Opt. Expr.}\ }\textbf {\bibinfo {volume}
  {14}},\ \bibinfo {pages} {8827} (\bibinfo {year} {2006})}\BibitemShut
  {NoStop}%
\bibitem [{\citenamefont {Schnell}\ \emph {et~al.}(2009)\citenamefont
  {Schnell}, \citenamefont {Garcia-Etxarri}, \citenamefont {Huber},
  \citenamefont {Crozier}, \citenamefont {Aizpurua},\ and\ \citenamefont
  {Hillenbrand}}]{Schnell2009}%
  \BibitemOpen
  \bibfield  {author} {\bibinfo {author} {\bibfnamefont {M.}~\bibnamefont
  {Schnell}}, \bibinfo {author} {\bibfnamefont {A.}~\bibnamefont
  {Garcia-Etxarri}}, \bibinfo {author} {\bibfnamefont {A.~J.}\ \bibnamefont
  {Huber}}, \bibinfo {author} {\bibfnamefont {K.}~\bibnamefont {Crozier}},
  \bibinfo {author} {\bibfnamefont {J.}~\bibnamefont {Aizpurua}}, \ and\
  \bibinfo {author} {\bibfnamefont {R.}~\bibnamefont {Hillenbrand}},\ }\href
  {\doibase 10.1038/NPHOTON.2009.46} {\bibfield  {journal} {\bibinfo  {journal}
  {Nat. Photon.}\ }\textbf {\bibinfo {volume} {3}},\ \bibinfo {pages} {287}
  (\bibinfo {year} {2009})}\BibitemShut {NoStop}%
\end{thebibliography}
\end{document}


\title{Evolutionary optimization of plasmonic nanostructures\\-- Supplementary material --}
\author{Feichtner Thorsten}
\author{Oleg Selig}
\author{Markus Kiunke}
\author{Bert Hecht}
\affiliation{Nano-Optics \& Biophotonics Group, Department of Experimental Physics 5, R\"ontgen
Research Center for Complex Material Research (RCCM), Physics Institute,
University of W\"urzburg, Am Hubland, D-97074 W\"urzburg, Germany}

\maketitle

\section{Size relation of matrix antenna and focal area}

\begin{figure}[htp]
	\centering
		\includegraphics[width=0.4\columnwidth]{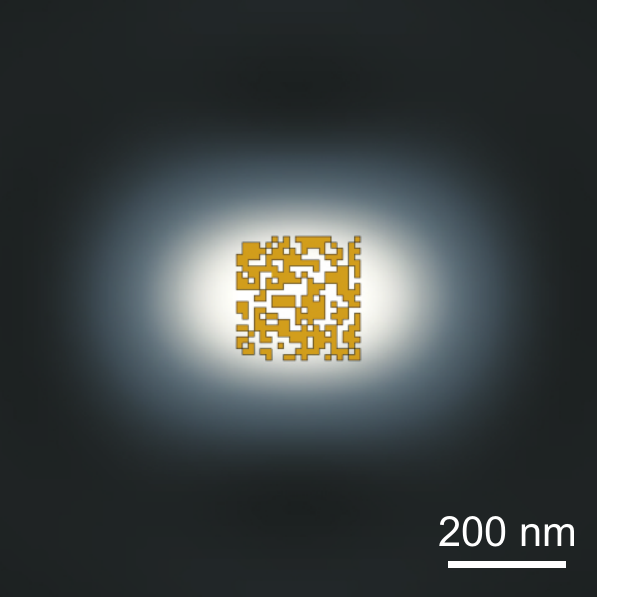}
	\caption{\label{fig:Focus_vs_matrix} Extension of the 21x21-matrix compared to the field profile of the 647 nm gaussian focus with $NA=1$.}
\end{figure}

The size of the matrix and the gold cubes were chosen such that the whole structure is illuminated by the excitation focus (Fig. S\ref{fig:Focus_vs_matrix}) and every cube is subjected to about the same light intensity.

\section{Number of non redundant binary odd-sized $n\times n$-Matrices with one symmetric and one antisymmetric axis imposed by the excitation polarization}
We consider a binary $n\times n$-matrix $A$ with odd $n$, $A_{x,y}\in [0,1]$ and $A_{(n+1)/2,(n+1)/2}=0$. It describes a plasmonic nano structure which is oriented in the $x$-$y$-plane centered on the $z$-axis and thus to the optical axis of the system. The linear excitation polarization is oriented along the $x$-axis resulting in electric fields, that are symmetric with respect to the $x$-$z$-plane and antisymmetric with respect to the $y$-$z$-plane. Consequently, structures that differ among each others only by a respective symmetry operation will result in the same physical behaviour, although the matrix apparently changes. For matrix sizes with $n\to\infty$ one can estimate that about 3/4 of the possible structures are physically redundant. Considerable computational effort can therefore be avoided by checking each newly generated structure for redundancy with already simulated structures.

First we calculate the number of structures with special symmetries within the set of all matrices as defined above.

\begin{figure*}[htp]%
\centering
\includegraphics[]{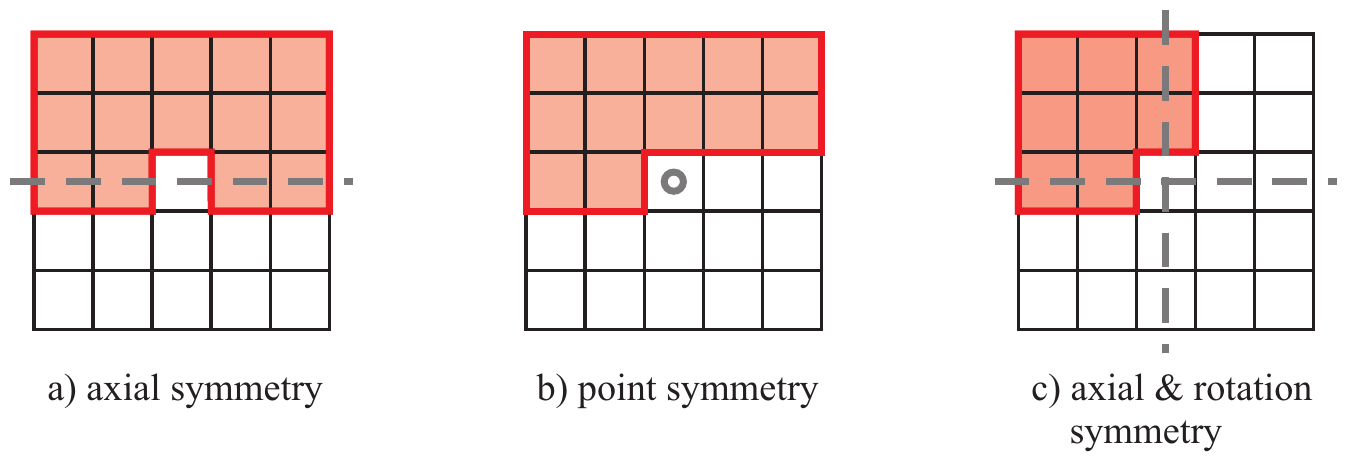}%
\caption{\label{fig:symms} At the example of $5\times5$ matrices, the necessary areas to describe structures with given symmetry axes (grey) are marked in red.}%
\end{figure*}

\subsection*{Axial symmetry}
In the case of axial symmetry the values in one row containing the center ($n-1$ elements) and a half matrix (rectangle of $n \cdot (n-1)/2$ elements; see fig. \ref{fig:symms}a) ) completely define a structure. The overall number $N_A$ of independent structures is thus:
\begin{equation}
N_A = 2^{(n-1) + n \cdot \frac{n-1}{2}} = 2^{(n-1)\left(\frac{n}{2} +1\right)}
\label{eq:achssym}
\end{equation}

\subsection*{Point symmetry}
In the case of point symmetry, the values in half a row ($(n-1)/2$ elements) and half a matrix completely describe the structures (see fig.~\ref{fig:symms}b)~), yielding
\begin{equation}
N_P = 2^{\frac{n-1}{2} + \frac{n(n-1)}{2}} = 2^{\frac{n^2-1}{2}}
\label{eq:punktsym}
\end{equation}
different configurations.

\subsection*{Point and axial symmetry}
Every matrix, which is symmetric to both axis is also point symmetric, and every matrix which is symmetric to the origin and to one axis, is also symmetric to the other axis. Thus only one overlapping set of all three single symmetry sets has to be computed. It is defined by a quarter matrix including the on-axis elements and excluding the center block (see Fig. S1 (c)):
\begin{equation}
N_{AP}=2^{\left(\frac{n+1}{2}\right)^2-1}
\label{eq:APsym}
\end{equation}

\subsection*{Counting}

Each matrix, which does not have one of the described symmetries can be transformed into three physical identical matrices by mirroring at the $x$- and $y$-axis. Thus the number of non physically redundant non symmetric structures is their complete number $N_{NS}$  divided by four. $N_{NS}$ can be computed as the number of all possible structures $N_\infty = 2^{n^2-1}$ minus all symmetric structures. As there are two mirror axes, the number of axis symmetric matrices $N_A$ has to be subtracted twice and the number of point symmetric structures once. The set of both point- and axis-symmetric matrices has then been subtracted three times and is re-added 2 times therefore. This leads to:

\begin{equation}
N_{NS} = \frac{N_\infty - 2 N_A - N_P + 2 N_{AP}}{4}
\label{eq:ex}
\end{equation}

Each axis and point symmetric structure can be mirrored into exactly one physical redundant but geometrical different structure. So to get the number of physical unique matrices, their number has to be divided by two after subtracting the structures with both symmetries.

\begin{equation}
N_{A,P} = \frac{2\cdot N_A + N_P - 3\cdot N_{AP}}{2} 
\label{eq:AundP}
\end{equation}

The structures with axes and point symmetry have to be counted completely, since no symmetry operation changes them neither geometrically nor physically.

The resulting complete number $N$ of physically unique square matrices with odd side length $n$ is:
\begin{align}
N & = N_{NS} + N_{A,P} + N_{AP} \\
\Rightarrow N &= \frac{1}{8} \left(2^{n^2}+2^{\frac{1}{2} \left(n^2+1\right)}+2^{\frac{1}{2}\left(n^2+n+2\right)}\right)
\label{eq:N}
\end{align}

\begin{figure*}[htp]%
\centering
\includegraphics[width=15cm]{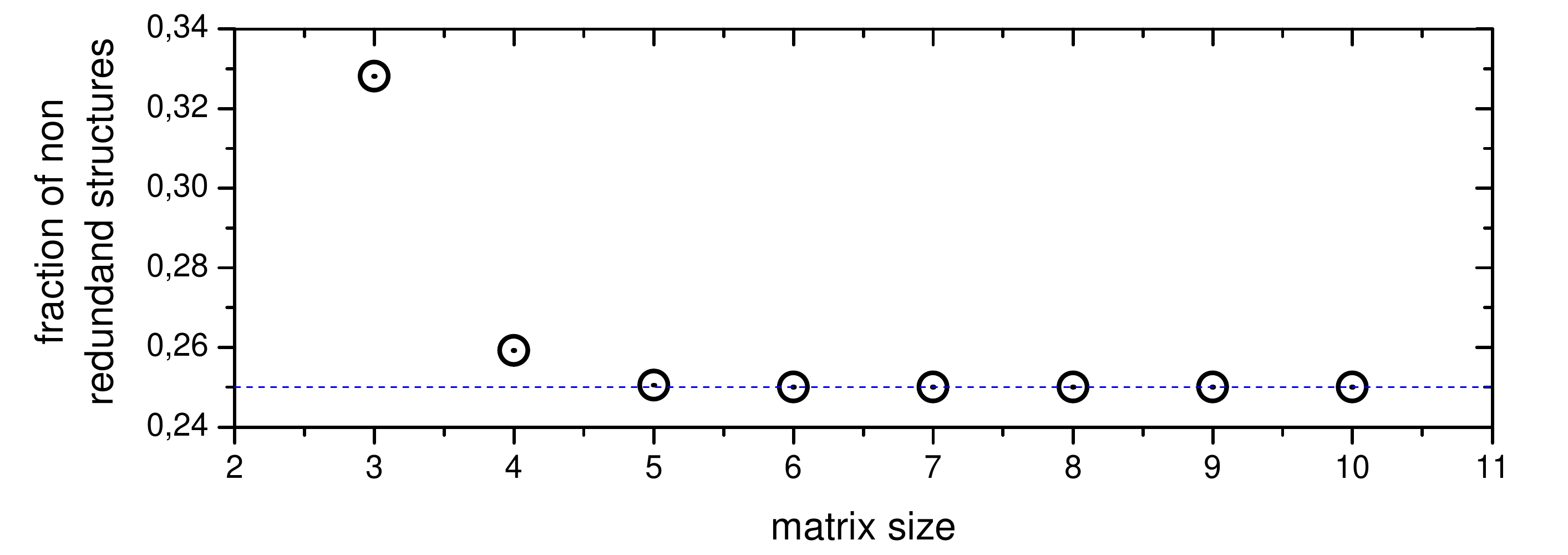}%
\caption{\label{fig:nonred} Amount of non redundant structures for the setup described in the text. For larger matrix sizes the limit of the fraction of not physically redundant square matrices is $0.25$ (blue dashed line).}%
\end{figure*}

An evaluation of this formula for increasing matrix sizes is depicted in Fig. S\ref{fig:nonred}. It shows, that for matrix sizes of 5 and bigger 75 percent of all structures are redundant. A considerable speed enhancement of the evolutionary algorithm is achieved by checking newly generated matrices for redundancies with already evaluated structures.

\section{Speed and Convergence}

To reduce the computational effort, simulations were terminated after 35 fs, when more than 98\% of the initial power has left the simulation volume. Although such a short simulation is not sufficient to reach a very high absolute accuracy, it is still possible to determine the relative hierarchy of simulated antennas in terms of fitness parameters such as near-field intensity. Considering that a full FDTD simulation of an antenna structure generally takes 70 -- 80~fs to fully converge, the overall simulation time is cut by about a factor of two. The best individuals are re-simulated until full convergence in order to obtain accurate results.

\section{Evolutionary algorithm}
\subsection{Methods of creating new individuals}

The EA uses different mechanisms to generate individuals for generation $n+1$ after evaluation of generation $n$:

\begin{figure*}[htp]
	\centering
		\includegraphics[width=0.5\columnwidth]{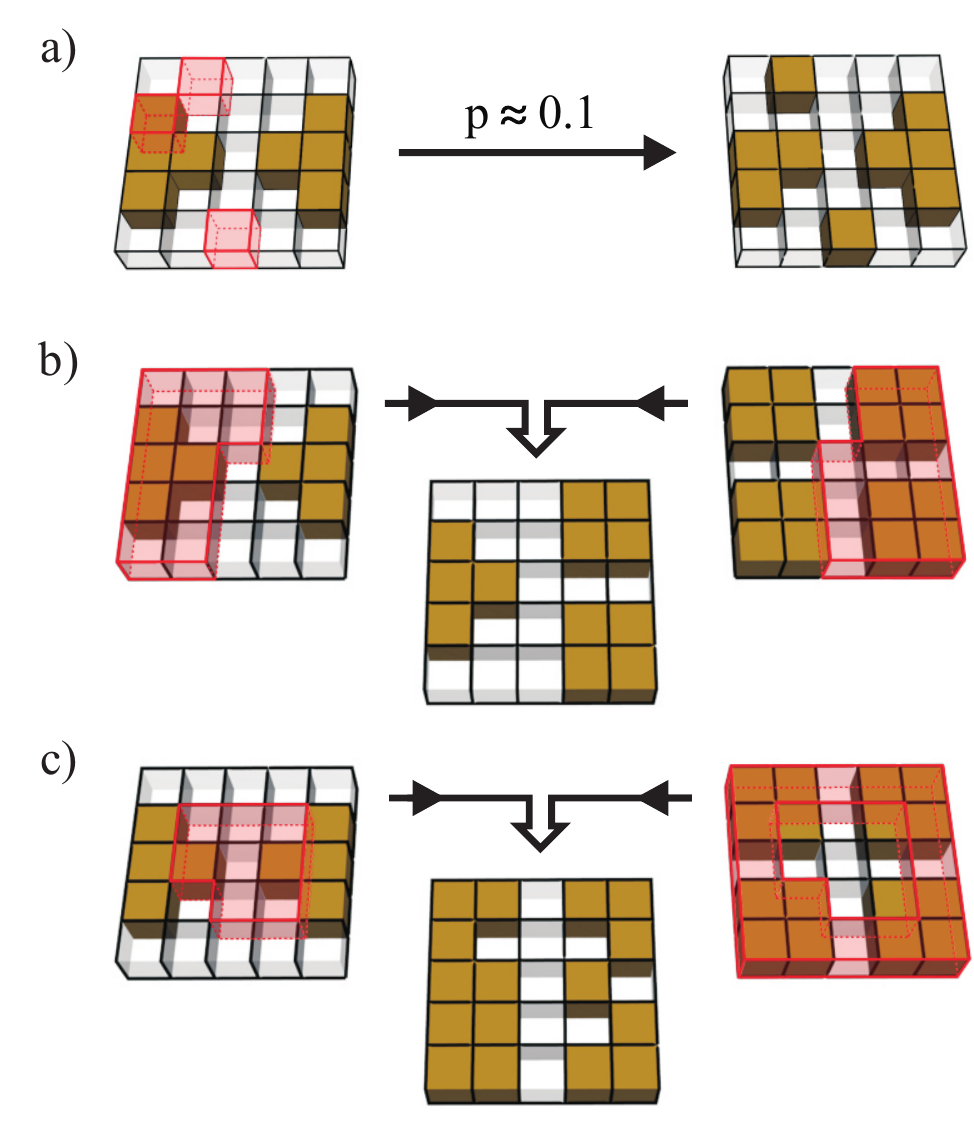}
	\caption{\label{fig:next_gen}  Illustration of methods used for creating descendants from eligible parents. (a) Mutation scheme. Each block is toggled with a certain probability. (b) and (c): Crossing of genomes after (b) column-by-column separation as well as (c) spiral-type separation.}
\end{figure*}

(i) \textit{Random:} Completely random structures are generated with a 50\% chance of each block being gold or void. These structures are independent of the parent pool and intended to introduce diversity into the genetic material.
	
(ii) \textit{Mutation:} Each block of parent $A$ is toggled between gold and void with a chance of 10\% (S4 (a)). This allows good structures to enhance further without intermixing with other genomes.

(iii) \textit{Linear and spiral crossing:} The line-by-line (spiral-) encoded  genomes [see S4 (b) and (c)] of two parents, $A$ and $B$, are crossed by combining the first half of the genome of $A$ with the second half of the genome of $B$ preserving the overall genome length \cite{Sivanandram2008}. The point at which the individual genomes are split is chosen randomly. This method allows the combination of the left (inner) part of parent antenna $A$ with the right (outer) part of a second parent $B$ [reddish areas in S4 (b) and (c)].

Each method to generate a subsequent individual is applied with equal probability. Method (iii) requires a second parent $B$ that also is chosen from the pool of the remaining 4 fittest individuals in the same way as $A$.  The new individuals are checked if they  or a physically redundant structure already has been simulated and replaced in that case with newly generated individuals (compare with S1).

\subsection{Development of higher-fitness individuals by means of evolutionary optimization}

\begin{figure*}[htp]
\includegraphics[width=0.7\textwidth]{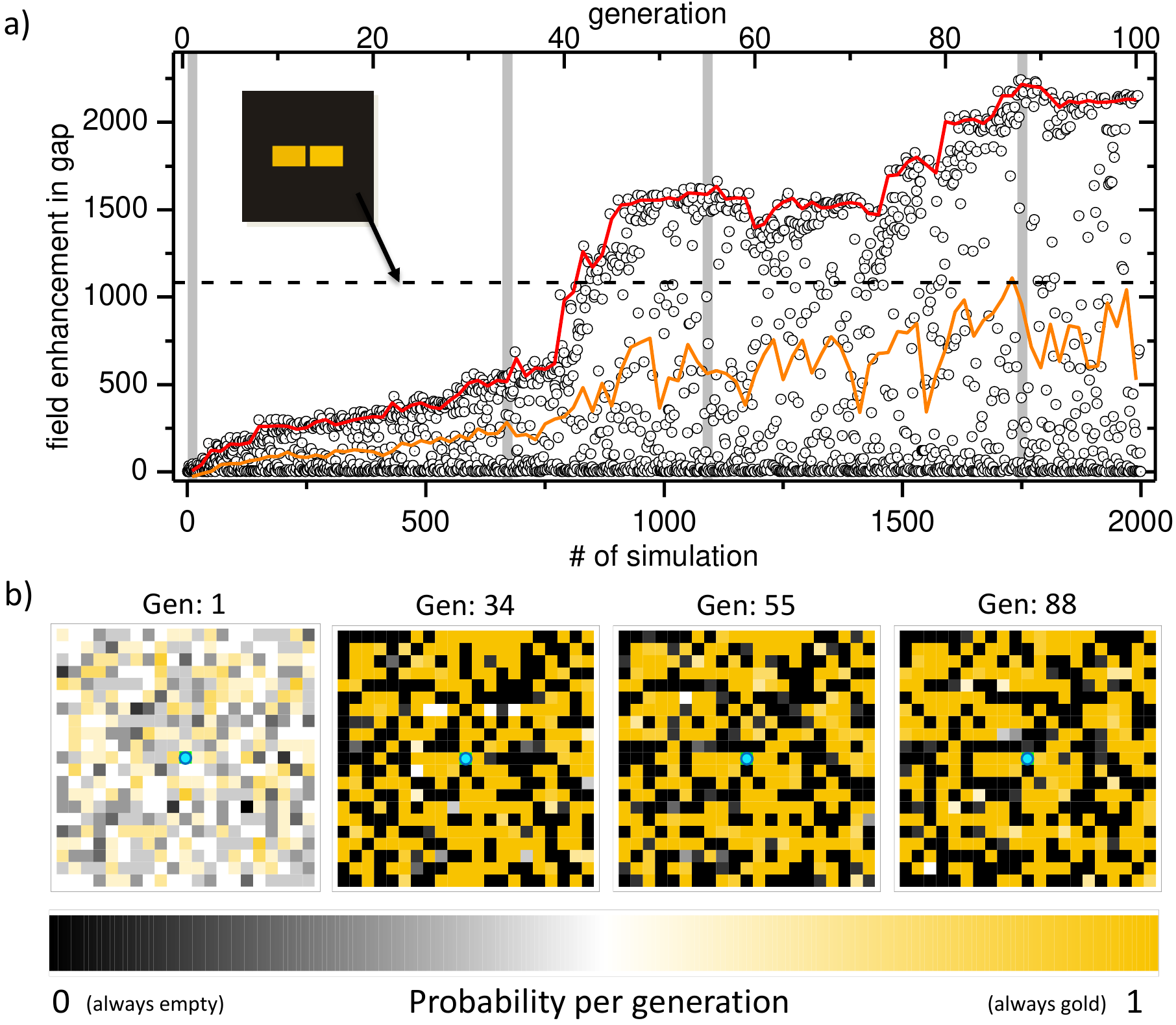}
\caption{\label{fig:propPerGen} Overview over the evolution process. a) shows the development of the near-field intensity enhancement as the number of generations progresses. Twenty simulations make up one generation. Number 1, 34, 55, and 88 are denoted by grey bars for further discussion in b). The red curve shows the maximal obtained enhancement, the orange the mean enhancement of a generation respectively. The black dashed horizontal line marks the field enhancement of the reference dipolar antenna. b) shows the probability of each 21$\times$21-matrix position for hosting a gold cube for the 10 best structures of the respective generation. The blue spot at the center marks the "gap", where the optimization took place and thus always was empty.}
\end{figure*}

Figure S\ref{fig:propPerGen} displays the progress of the evolutionary optimization as a function of the number of generations. Fig. S\ref{fig:propPerGen} (a) shows the near-field intensity enhancement in the gap of all 2000 simulated individuals of the first run. It is important to note, that the individuals have only been simulated for 35 fs and therefore their fitness parameter is underestimated typically by 10--20\%. The first generation contained 20 random individuals with field enhancements between 0.2 and 43.8. Figure \ref{fig:propPerGen} (b) shows, to the very left, the probability of each matrix space of the best 10 structures of the first generation to be occupied with gold (Gen: 1). The low saturation reflects the randomness of each single matrix position. While the algorithm develops, the near-field intensity enhancement gradually increases and surpasses the enhancement of the reference antenna in generation 42.

The probability plots in Fig. \ref{fig:propPerGen} (b) clearly illustrate the development of some structural features: left of the gap a small arm with 40$\times$20 nm size develops, while on the right side a more triangular structure emerges. On the way to a better structure the individuals in a single generation get more similar within the first 10 generations and show only small differences between the best single structures henceforth. The development of some important features leads to a jump in the maximum fitness of a generation, when they are established the first time. The first such feature is a gap, which ensures the minimal possible distance of charge carriers to the optimization position. Then the areas next to the gap are getting more solid, to establish currents and supply the gap with a higher surface charge density. The first big jump in the fitness originates from the removal of the block directly above the center, opening the gap.

The fitness also occasionally reduces which may occur, when a local maximum of the configuration space was found and among the offspring structures there is not yet an individual that includes a new key feature. More and more individuals will then be
generated by mutation, thus further lowering the fitness in the beginning but also generating new features, which will finally lead to the formation of new structural elements that will then generate better structures.

\section{Quality factor}

The optimal antenna structure found by the evolutionary algorithm exhibits a comparatively large quality factor $Q \approx 20$. Due to the effective wavelength scaling \cite{Novotny2007} the resonance length of plasmonic two-wire antennas strongly depends on the wire cross-section. A small cross section $d$ leads to a reduced overall length for resonant antennas which in turn leads to a reduced radiation resistance. The reduced radiation resistance directly results in an increased quality factor of the antenna resonance. Table \ref{Tab:AsRa} demonstrates this behaviour for a series of antennas with the same resonance position, but varying cross section $d$.  

\begin{table}[htp]%
\includegraphics[scale=0.75]{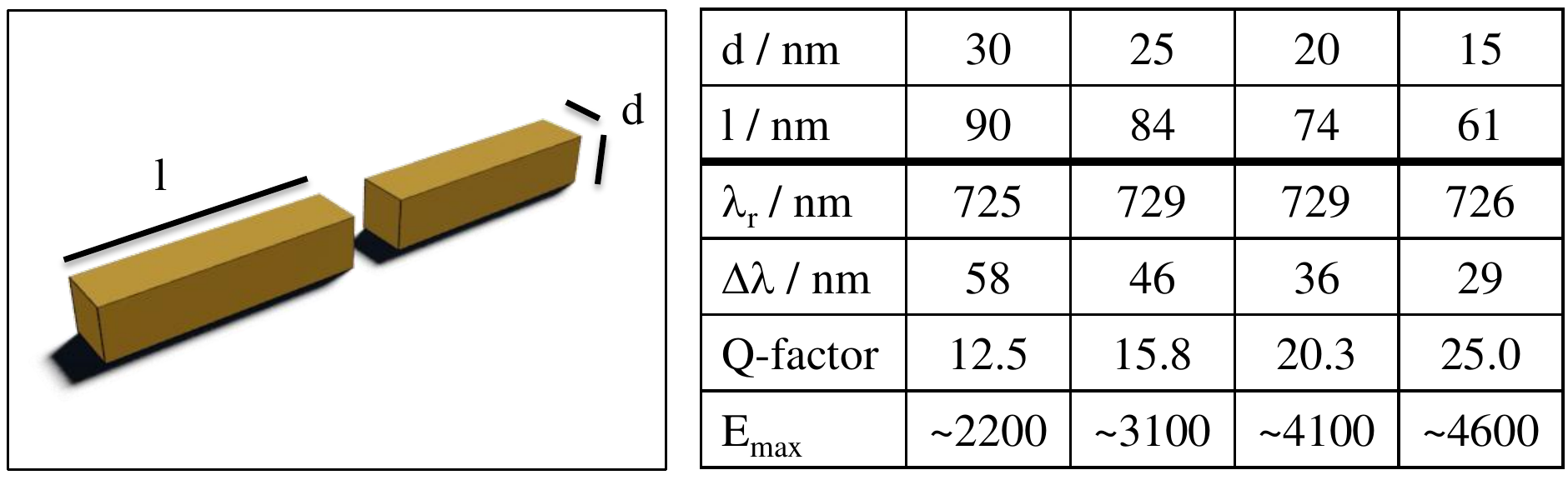}%
\caption{Q-factor, resonance width $\Delta \lambda$, and maximum near-field intensity enhancement $E_\text{max}$ of optical antennas consisting of gold wires with length $l$ separated by a 10 nm gap with constant resonance wavelength $\lambda_\text{r}$ but decreasing square cross section $d^2$.}%
\label{Tab:AsRa}%
\end{table}

Along with the higher Q-factors, thinner antennas also show higher maximum near-field intensity enhancement. This is due to the better field confinement in the gap due to the smaller cross section and thus the enhanced lightning rod effect.

\section{Radiation pattern}
\begin{figure}[htp]
	\centering
		\includegraphics[scale=0.75]{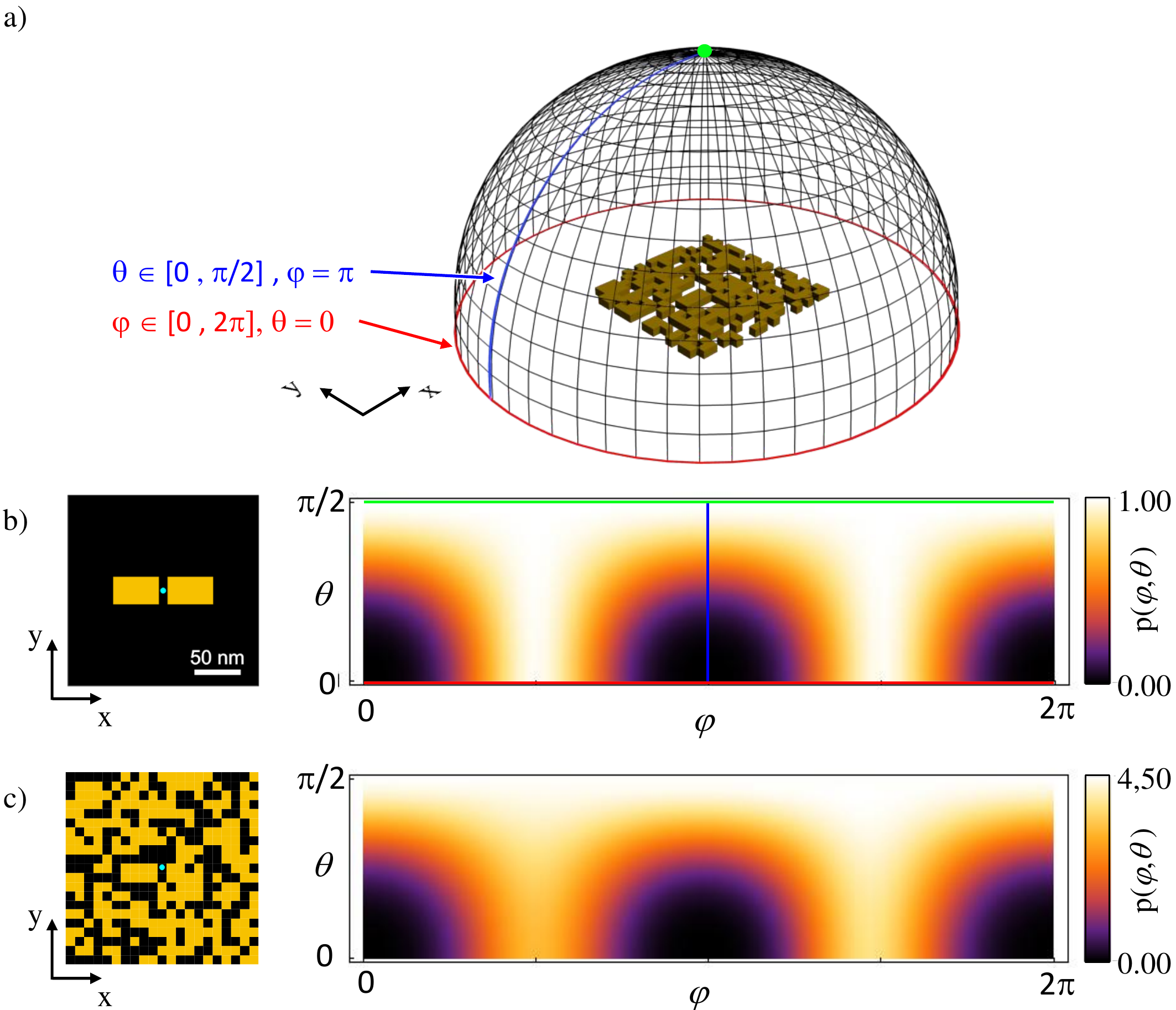}
	\caption{\label{fig:FarFieldSketch} Radiation patterns of reference and fittest matrix antenna. a) Definition of spherical coordinates $\theta$ and $\varphi$ with respect to the antenna orientation. A point dipole emitter emitting at the resonance wavelength $\lambda_\text{res}=650$ nm is placed at the position of highest near-field intensity enhancement. Since the structures are symmetrical in $z$-direction, only the radiation power in the upper half space needs to be evaluated. The coloured markings (red, blue, green) denote particular constant-coordinate lines. b) shows geometry and normalized dipolar radiation pattern in cylindrical projection for the reference antenna. c) shows geometry and radiation pattern of the fittest matrix antenna normalized to the reference antenna.}
\end{figure}

To demonstrate the validity of the reciprocity theorem for the fittest antenna found by the evolutionary algorithm, we place a dipole emitter at the position of highest near-field intensity enhancement and study its enhanced emission. The resulting far-field power flow is depicted in Fig. \ref{fig:FarFieldSketch}.

The directivity of an antenna is defined as the maximum of the directive gain
\begin{equation}
D(\theta,\varphi) = \frac{4\pi \cdot p(\theta,\varphi)}{P_\text{rad}}
\label{eq:}
\end{equation}
with $P_\text{rad}$ the integrated radiated power computed from the integral over the radiation power per unit solid angle $p(\theta,\varphi)$. The reference antenna shows a symmetrical dipolar radiation pattern with a directivity of $D_\text{ref}=2.09$ dBi which is only slightly smaller than $D_{\lambda/2} = 2.15$ dBi of an ideal thin wire half wave dipole.

The radiation pattern of the best matrix antenna deviates only slightly from reference antenna, but exhibits a more pronounced radiation in the $z$-direction. Considering reciprocity, its enhanced directivity of $D_\text{EA} = 2.31$ dBi reflects the optimization to irradiation by a gaussian beam.